\documentclass[journal]{IEEEtran}

\usepackage{graphicx}
\usepackage{amsmath,amssymb,amsfonts}
\usepackage{multirow}
\usepackage{psfrag}
\usepackage{subfigure}
\usepackage{color}
\usepackage{soul}
\usepackage{xcolor}
\usepackage[hyphens]{url}
\usepackage[bookmarks=false]{hyperref}
\usepackage[hyphenbreaks]{breakurl}
\usepackage{lipsum}
\usepackage[ruled,vlined,linesnumbered]{algorithm2e}
\usepackage{mathrsfs}
\usepackage{pbox}
\usepackage{amsfonts}
\usepackage{mathtools}
\usepackage{amsthm}
\usepackage{pbox}
\usepackage{wrapfig}
\usepackage{array}
\usepackage[utf8]{inputenc}
\usepackage{longtable}
\usepackage{tikz}

\def\checkmark{\tikz\fill[scale=0.4](0,.35) -- (.25,0) -- (1,.7) -- (.25,.15) -- cycle;}

\DontPrintSemicolon

\newcommand{\matr}[1]{\mathbf{#1}}

\theoremstyle{definition}
\newtheorem{definition}{Definition}[section]

\DeclareMathOperator*{\argmaxA}{arg\,max}

\makeatletter
\def\hlinewd#1{%
  \noalign{\ifnum0=`}\fi\hrule \@height #1 \futurelet
   \reserved@a\@xhline}
\makeatother

\graphicspath{{Figures/}}
\ifCLASSINFOpdf
\else
\fi

\begin{document}

\title{Reducing Operation Cost of LPWAN Roadside Sensors Using Cross Technology Communication}

\author{Navid~Mohammad~Imran and Myounggyu~Won\\
Department of Computer Science\\
University of Memphis, TN, United States\\
\{nimran, mwon\}@memphis.edu
}

\markboth{Journal of \LaTeX\ Class Files,~Vol.~13, No.~9, September~2014}%
{Shell \MakeLowercase{\textit{et al.}}: Bare Demo of IEEEtran.cls for Journals}

\maketitle

\begin{abstract}
Low-Power Wide-Area Network (LPWAN) is an emerging communication standard for Internet of Things (IoT) that has strong potential to support connectivity of a large number of roadside sensors with an extremely long communication range. However, the high operation cost to manage such a large-scale roadside sensor network remains as a significant challenge. In this article, we propose Low Operation-Cost LPWAN (LOC-LPWAN), a novel optimization framework that is designed to reduce the operation cost using the cross-technology communication (CTC). LOC-LPWAN allows roadside sensors to offload sensor data to passing vehicles that in turn forward the data to a LPWAN server using CTC aiming to reduce the data subscription cost. LOC-LPWAN finds the optimal communication schedule between sensors and vehicles to maximize the throughput given an available budget. Furthermore, LOC-LPWAN optimizes the fairness among sensors by preventing certain sensors from dominating the channel for data transmission. LOC-LPWAN can also be configured to ensure that data packets are received within a specific time bound. Extensive numerical analysis performed with real-world taxi data consisting of 40 vehicles with 24-hour trajectories demonstrate that LOC-LPWAN reduces the cost by 50\% compared with the baseline approach where no vehicle is used to relay packets. The results also show that LOC-LPWAN improves the throughput by 72.6\%, enhances the fairness by 65.7\%, and reduces the delay by 28.8\% compared with a greedy algorithm given the same amount of budget.
\end{abstract}


\IEEEpeerreviewmaketitle

\section{Introduction}
\label{sec:introduction}

Low-Power Wide-Area Network (LPWAN) is an emerging network paradigm for Internet of Things (IoT) designed to support the connectivity of a huge number of low-power wireless devices leveraging extremely long-range communication at a low cost~\cite{li2018lpwan}\cite{patel2017experimental}. With growing demand for the LPWAN technology, numerous LPWAN standards have been developed such as Narrowband-IoT (NB-IoT)~\cite{hoglund2017overview}, Long-Term Evolution for Machines (LTE-M)~\cite{ltem}, 5G~\cite{ngmn}, Sensor Network over White Spaces (SNOW)~\cite{saifullah2016snow}\cite{saifullah2018low}, Long Range Wide Area Network (LoRa)~\cite{de2017lorawan}, SigFox~\cite{gomez2019sigfox}, \emph{etc.} While LPWAN is increasingly used in various IoT applications such as smart irrigation~\cite{zhao2017design}, smart agriculture~\cite{vuran2018internet}, smart health~\cite{petajajarvi2016evaluation}, and smart buildings~\cite{centenaro2016long}, the vehicular section has received noteworthy attention due to the significant potential of the LPWAN technology for connecting a large number of roadside sensors~\cite{santa2019lpwan}\cite{kombate2016internet}. As such, various LPWAN-based intelligent transportation systems (ITS) applications have been developed~\cite{de2020review}\cite{rahman2019implementation}, \emph{e.g.,} parking surveillance system~\cite{ke2020smart,jeon2018design}, vehicle diagnostic system~\cite{chou2017car}, and vehicular monitoring system~\cite{hsieh2017vehicle}.

A significant challenge for establishing connectivity of roadside sensors based on LPWAN is the high operation cost especially when licensed/cellular bands are used (\emph{e.g.,} NB-IoT~\cite{hoglund2017overview}, LTE-M~\cite{ltem}, 5G~\cite{ngmn}) that require subscription fees for data usage. At present, different kinds of data plans are available for both LTE-M and NB-IoT from the cellular network providers in the US (\emph{e.g.,} AT\&T, Verizon and T-Mobile). For LTE-M, AT\&T charges \$1 per month for one device with a 500 KB data limit~\cite{att}. Verizon, on the other hand, offers the same 500 KB data per month with a costlier \$3 package~\cite{verizon}. T-Mobile provides 1 MB data per month at the cost of \$0.5 for both LTE-M and NB-IoT~\cite{tmobile}. There are some approaches based on non-licensed/non-cellular network (\emph{e.g.,} SNOW~\cite{saifullah2016snow}\cite{saifullah2018low}, LoRa~\cite{de2017lorawan}, SigFox~\cite{gomez2019sigfox}) that can be used to save the cost. However, when covering a large area of interest with a massive number of sensors, the cost for such LPWAN technologies could possibly be higher than the subscription-based LPWAN due to the high operation and maintenance cost. More specifically, creating a large-scale non-cellular LPWAN requires very carefully designed node deployment plans and frequent maintenance and monitoring of the network due to the fluctuating channel conditions and relatively shorter communication range especially in urban environments (\emph{e.g.,} 5 km~\cite{buurman2020low}) because of significant obstructions~\cite{buurman2020low}.

To address the cost issue, numerous approaches have been attempted~\cite{gu2020survey}. We categorize these solutions largely into three groups: (1) hardware-based, (2) data reduction-based, and (3) free channel-based solutions. The hardware-based methods basically reduce the device cost by developing a LPWAN node with low-cost hardware design, \emph{e.g.,} by developing a low-cost radio interface~\cite{thoen2019deployable} or adopting a low-cost processor~\cite{sallouha2017ulora}. However, these hardware-based cost reduction methods are effective only for the initial deployment stage, failing to lower the major cost for the monthly or yearly subscription fees. The data reduction-based methods focus on algorithm design to reduce the size of data transmitted by LPWAN nodes, \emph{e.g.,} through dynamic re-sampling~\cite{botero2018data} and sensor data integration~\cite{silva2019low}. Although these methods reduce the data size effectively to decrease the cost, the data cost cannot be removed entirely as LPWAN nodes still transmit sensor data regardless of the data size. The free channel-based methods allow LPWAN nodes to communicate via unlicensed channels such as TV white spaces~\cite{saifullah2016snow}\cite{saifullah2018low} to completely remove the data cost. However, depending on the number of connected devices on a shared spectrum, relying on an unlicensed band may cause significant performance degradation~\cite{raza2017low}.

In this article, we propose LOC-LPWAN (Low Operation-Cost LPWAN), an incentive-based optimization framework designed to minimize the operation cost of roadside sensors using the cross-technology communication (CTC)~\cite{chen2020reliable}. It achieves significant cost savings by allowing LPWAN sensors to offload packets to vehicles using CTC that enables communication between devices equipped with heterogeneous communication technologies. More precisely, LPWAN sensors transmit sensor data to vehicles equipped with 5G/LTE/Dedicated Short-Range Communications (DSRC)-based Vehicle-to-everything (V2X) communication modules which in turn forward the data to a LPWAN server, thereby saving the cost for using the licensed band for LPWAN for data transmission. LOC-LPWAN is designed to compute the optimal data communication schedule between LPWAN sensors and vehicles such that the total amount of data transmitted by sensors is maximized given the limited budget and the minimum amount of compensation paid to participating vehicles. The communication schedule is derived based on the regular trajectories of vehicles such as bus routes and commuting routes. LOC-LPWAN is capable of adapting the communication schedule in response to dynamically changing traffic conditions, allowing the user to maximize the benefits within a budget constraint. Additionally, LOC-LPWAN can be configured to achieve the maximum degree of fairness among sensors in terms of how much data each sensor transmits, \emph{i.e.,} preventing a certain sensor from dominating the opportunity to transmit data. Furthermore, it allows for delay-bound data transmission, \emph{i.e.,} ensuring that sensor data are received within a specified delay bound.

An extensive numerical study is conducted to evaluate the performance of LOC-LPWAN using the T-Drive trajectory dataset~\cite{yuan2011driving}\cite{yuan2010t} which comprises of one-week GPS trajectories of taxis in Beijing, China. The performance of LOC-LPWAN is compared with a greedy approach where sensors transmit data whenever there is a vehicle in its communication range as long as the available budget is not exhausted. The results demonstrate that LOC-LPWAN identifies the optimal communication schedule and participating vehicles to significantly increase the total amount of data transmitted. It is also shown that LOC-LPWAN successfully achieves fair and delay-bound data transmissions. Specifically, LOC-LPWAN improves the throughput by 72.6\%, enhances the fairness of data transmission by 65.7\%, and reduces the delay between the sensor data generation time and the data transmission time by 28.8\%, in comparison with a greedy algorithm. The performance of LOC-LPWAN is also compared with the baseline algorithm where no vehicle is used for relaying packets. The results demonstrate that LOC-LPWAN reduces the cost by 50\% compared with the baseline approach. Overall, our contributions are summarized as follows.

\begin{itemize}
  \item A novel cross-technology communication (CTC)-based approach is proposed to reduce the operation cost of LPWAN-based roadside sensors.
  \item A novel optimization framework is designed to find the optimal communication schedule between LPWAN sensors and vehicles given budget constraints.
  \item The proposed optimization framework achieves fair data transmission for LPWAN sensors as well as delay-bound data transmission that guarantees sensor data to be transmitted within a specific time bound.
  \item An extensive numerical study using the real-world global positioning system (GPS) trajectories of taxis is conducted to demonstrate the effectiveness of the proposed optimization framework.
\end{itemize}

This paper is organized as follows. In Section~\ref{sec:related_work}, we review the related work focusing on the cost reduction mechanisms for LPWAN as well as the cross-technology communication. The system model and notations used throughout this article are defined in Section~\ref{sec:system_model}. Based on this system model, we present the details of LOC-LPWAN in Section~\ref{sec:proposed_approach} in which we discuss the optimal scheduling for communication, fair and delay-bound data transmissions for sensors. Simulation results are presented in Section~\ref{sec:evaluation}. Finally, several issues and limitations of LOC-LPWAN are discussed in Section~\ref{sec:discussion}, and we conclude in Section~\ref{sec:conclusion}.

\section{Related Work}
\label{sec:related_work}


\subsection{Cost Reduction Mechanisms for LPWAN}
\label{subsec:cost_reduction}

Despite the promising properties of LPWAN, the high operation cost for a large number of LPWAN nodes is a major hurdle for industry and academia to adopt the technology~\cite{sallouha2017ulora}. To address this cost issue, numerous cost reduction mechanisms have been developed. We categorize these works into three different groups: hardware-based, data reduction-based, and free channel-based methods.

\textbf{Hardware-Based Methods:} These approaches reduce the cost by designing more effective hardware for LPWAN nodes. For example, the device cost is reduced by developing a power-efficient radio interface~\cite{thoen2019deployable}. The power efficiency is derived by controlling the lossy power circuitry on the application microcontroller. With this new hardware design, up to 20 years of operation time can be achieved for a spreading factor of 7~\cite{thoen2019deployable}. A novel LPWAN platform, uLoRa, is designed using a ultra low-power ARM processor to reduce the device cost~\cite{sallouha2017ulora}. The device cost is further reduced by implementing a spreading factor adaptation algorithm and improved peripheral and timer configuration~\cite{sallouha2017ulora}. A novel multiplier-less filter is developed based on the chirp segmentation and quantization to reduce the cost~\cite{stewart2019reducing}. Use of both methods exponentially decrease the number of required input samples for the multiplier-less filter, thus reducing the cost. It is also shown the decoding performance of LoRa devices remains stable for moderate quantization which allows the use of higher quantization factors to compensate for added complexity~\cite{stewart2019reducing}. A low-cost testbed, Chirpbox is developed to support large-scale deployment of LPWAN nodes~\cite{ma2020poster}. Another cost-effective testbed based on power-supply and power-management board is developed~\cite{pawar2020low}. The power-management is achieved by a technique called power gating where the power is briefly cut off from the circuit when not needed and turned back on when it is needed~\cite{pawar2020low}. However, although these hardware-based mechanisms reduce the initial deployment cost by lowering the device cost, they are incapable of cutting down the monthly or annual data subscription fees.

\textbf{Data Reduction-Based Methods:} These methods are designed to reduce the cost by decreasing the data packet size. Botero \emph{et al.} reduce the packet size through dynamic subsampling, data fusion, and data scaling~\cite{botero2018data}. The dynamic sampling is achieved using 1-D Kalman filter, data fusion and data scaling by removing unimportant decimal places~\cite{botero2018data}. A novel data integration method is developed to reduce the size of packets~\cite{silva2019low}. A cost analysis on varying sensing intervals is performed~\cite{sherazi2018energy}. In particular, it is shown that in the presence of harvested energy, the battery replacement cost is significantly decreased which drops the sensing interval of Long Range Wide Area Network (LoRaWAN) sensors and their operational cost~{\cite{sherazi2018energy}}. Although the reduced packet size contributes to lowering the cost, these methods still require LPWAN nodes to transmit packets to the server, thereby failing to remove the data cost completely. In contrast, our approach attempts to eliminate the data cost by allowing vehicles to offload their packets to passing vehicles via CTC. In particular, we note that these data reduction-based methods can be very useful when combined with our approach since LPWAN nodes can transmit more packets to vehicles due to the smaller packet size.

\textbf{Free Channel-Based Methods:} These mechanisms are designed to reduce the cost by exploiting unlicensed bands for transmitting data. A novel LPWAN technology that operates over the TV white spaces called SNOW is developed~\cite{saifullah2016snow}\cite{saifullah2018low}. The distributed OFDM (Orthogonal Frequency Division Multiplexing) based technology allows LPWAN communication with a single antenna and a simple physical layer design which reduces both the deployment and operational cost~{\cite{saifullah2016snow}}{\cite{saifullah2018low}}. To reduce the high device cost of SNOW (\emph{i.e.,} Universal Software Radio Peripheral (USRP) devices being used as LPWAN nodes) and the large form factor of the LPWAN nodes designed for unlicensed bands, a low-cost, low-power, and low form-factor LPWAN nodes based on the TV white space are also developed~\cite{rahman2019implementation}\cite{ismail2018demo}. These approaches decrease the cost and form-factor by 25x and 10x. The cost reduction comes from using commercial off-the-shelf components for SNOW implementation. Challenges related to reliability and range due to this implementation are also addressed by proposing different methods such as adaptive transmission power control and allocating special subcarriers~\cite{rahman2019implementation}\cite{ismail2018demo}. However, such free channels are not always available; as such, depending on the demand of connected devices~\cite{raza2017low}, the performance can be degraded. Currently this demand is partly hampered by the difficulty of providing LPWA as a service to consumers. In addition, interoperability between different LPWA technologies is still not explored deeply and security vulnerabilities still need to be addressed~\cite{raza2017low}. Table I presents a summary of the characteristics of the LPWAN cost reduction methods.

\begin{table}[h]
\center
\label{table:notations}
\caption{LPWAN Cost Reduction Methods}
\begin{tabular}{|m{0.6cm}|m{1cm}|m{1cm}|m{1cm}|m{1cm}|m{1cm}|}
\hline
Papers & Device cost & Data Packet size & Network Band & Date Rate & Scalability \\
\hline
\cite{thoen2019deployable} & \checkmark &  &  & \checkmark & \\
\hline
\cite{sallouha2017ulora} & \checkmark &  &  &  & \checkmark\\
\hline
\cite{botero2018data} &  & \checkmark &  &  & \checkmark\\
\hline
\cite{silva2019low} &  & \checkmark &  & \checkmark & \\
\hline
\cite{raza2017low} &  &  & \checkmark &  & \checkmark\\
\hline
\cite{stewart2019reducing} & \checkmark &  &  &  & \\
\hline
\cite{ma2020poster} & \checkmark &  &  &  & \checkmark\\
\hline
\cite{pawar2020low} & \checkmark &  &  &  & \\
\hline
\cite{sherazi2018energy} &  & \checkmark &  &  & \\
\hline
\cite{ismail2018demo} &  &  & \checkmark &  & \\
\hline
\cite{saifullah2016snow} &  &  & \checkmark & \checkmark & \checkmark\\
\hline
\cite{saifullah2018low} &  &  & \checkmark & \checkmark & \checkmark\\
\hline
\cite{rahman2019implementation} & \checkmark &  & \checkmark &  & \\
\hline
\end{tabular}
\end{table}

\subsection{Cross Technology Communication (CTC)}
\label{subsec:cross_platform}

Cross-Technology Communication (CTC) enables direct communication among devices running different communication technologies~\cite{liu2019lte2b}. A signal emulation technique is developed to use Wi-Fi signals to emulate ZigBee signals~\cite{chen2020reliable}. A symbol-level encoding mechanism is developed to enable more efficient ZigBee to Wi-Fi CTC~\cite{wang2018symbol}. BLE2LoRa enables CTC between Bluetooth Low Energy (BLE) and LoRaWAN (LPWAN) using the frequency shifting of BLE devices~\cite{lible2lora}. LoRaBee is another solution that enables CTC between ZigBee and LoRa (LPWAN)~\cite{shi2019lorabee}.

Numerous research works have been conducted to enhance the effectiveness of CTC. Transmitter-transparent CTC is developed to significantly improve the throughput of CTC by leaving the processing complexity completely at the receiver side~\cite{guo2019lego}. B$^2$W$^2$ is developed to enable N-way concurrent communication among devices with different communication technologies (\emph{i.e.,} Wi-Fi, and BLE)~\cite{chi2019concurrent}. To overcome the high quantization errors due to the frequency domain-based approaches and improve reliability, the first time-domain emulation (TDE) method is developed~\cite{liu2019lte2b}. TwinBee is developed to recover the intrinsic emulation errors~\cite{chen2020reliable}. Another method to improve the reliability through bidirectional CTC design is developed~\cite{he2020reliable}. A routing protocol for CTC is developed to allow packets routed among devices with heterogeneous communication technologies~\cite{gao2019rowbee}. In this work, we assume that CTC is used to enable direct communication between LPWAN nodes and vehicles that use different communication technologies, \emph{i.e.,} LPWAN, and 5G/LTE/DSRC-based V2X, respectively.

\subsection{Sensor Data Collection Using Vehicles/UAVs}
\label{subsec:sensor_data_collection}

As IoT applications covering a wide area are increasing, establishing and maintaining connectivity of IoT sensors to an infrastructure network have become ever more challenging. A possible approach to address this challenge is to utilize mobile data collectors such as vehicles and UAVs~\cite{li2021drlr}. Numerous approaches based on mobile data collectors have been proposed focusing on various aspects such as data security~\cite{huang2020intelligent}\cite{ouyang2021verifiable}\cite{jiang2020trust}, wireless communication between sensors and vehicles~\cite{nnamani2021joint}, energy efficiency of data collection~\cite{fu2021energy}, and operation cost~\cite{li2021drlr}.

There are some solutions designed to ensure data security on collected sensor data. Huang \emph{et al.} aim to minimize the false ratio
(\emph{i.e.,} the ratio of bogus packets to total packets) and packet dropping rate (\emph{i.e.,} deliberately not presenting data)~\cite{huang2020intelligent}. Ouyang \emph{et al.} study the trust verification problem for a mobile data collector (UAV) to counter false and malicious data~\cite{ouyang2021verifiable}. Shen \emph{et al.} develop a data collection method to collect data from trusted sensors~\cite{shen2021attdc}. Jiang \emph{et al.} perform a similar study as~\cite{ouyang2021verifiable} but aim to maximize the
network lifetime by effectively constructing the UAV trajectories~\cite{jiang2020trust}. Nnamani \emph{et al.} also focus on the security of data collection but put more emphasis on the secure communication between UAVs and sensors~\cite{nnamani2021joint}. Compared to these solutions, our approach is focused on determining the optimal communication schedule to minimize the operation cost.

Various other aspects of the sensor data collection problem have been investigated. Fu \emph{et al.} study the data collection problem using UAVs specifically concentrating on the limited battery capacity of UAVs~\cite{fu2021energy}. A method to plan a route of UAVs is designed such that the battery of UAVs can be recharged via the wireless charging station at the center of a grid
where an area of interest is represented as a set of grids~\cite{fu2021energy}. Qin \emph{et al.} address the problem of ensuring the timeliness of sensor data~\cite{qin2021distributed}. Our proposed framework also provides an option for the user to enable delay-bound sensor data collection, our work is different because Qin \emph{et al.} assume that vehicles are the ones that generate sensor data.

Li \emph{et al.} propose a similar approach as LOC-LPWAN in that it is designed to reduce the operation cost~\cite{li2021drlr}. Their work is focused on the coverage of a vehicle, \emph{i.e.,} providing higher rewards to the vehicles if they cover a sensor that is not visited by a sufficient number of vehicles. However, they do not take into account the time factor, \emph{i.e.,} ``when'' a vehicle will visit sensors as the vehicle trajectory is not defined on a time-varying spatial domain, which is critical in determining the optimal communication schedule between sensors and vehicles. For example, even if a sensor is visited frequently by many vehicles in a short period of time, it may not be effective if the sensor has not generated much data during that short period of time. On the other hand, our optimization framework is defined on a time-varying spatial domain to incorporate the vehicle's driving schedule in determining the optimal communication schedule. Specifically, our work takes into account the time-varying distance between a vehicle and a sensor to adapt the communication schedule according to dynamically changing traffic conditions. Furthermore, in determining the optimal communication schedule, LOC-LPWAN incorporates numerous practical factors such as the sensor data generation rate ($\alpha_{rate}$), vehicular communication range ($\alpha_{range}$), and data transmission cost for each sensor ($\alpha_{cost}$). Additionally, our work provides options to users to enable fair and delay-bound sensor data collection along with cost reduction.

\section{System Model}
\label{sec:system_model}

We consider a two-dimensional target area. The time is
discretized into time slots denoted by $T = \{t_1, t_2, ..., t_n\}$. The duration of each time slot is 1 second. It is worth to note that the timescale can be easily changed to make it applicable to different applications. There are $N_v$ vehicles denoted by $V=\{v_1, v_2, ..., v_{N_v}\}$ that wish to participate in the incentive program (\emph{i.e.,} if selected, the vehicle will relay data packets received from sensors and earn money). Each vehicle $v \in V$ has a set of estimated trajectories denoted by $\mathbf{R_v} = \{R_v^1, R_v^2, ..., R_v^n\}$. An $i$-th trajectory of vehicle $v \in V$ denoted by $R_v^{i} \in \mathbf{R_v}$ is a sequence of GPS points, \emph{i.e.,} $R_v^{i} = \{r_{v,i}^1, r_{v,i}^2, ..., r_{v,i}^{|R_v^{i}|}\}$ where $r_{v,i}^j \in \mathbb{R}^2$. Corresponding to $\mathbf{R_v}$ is a set of time traces $\mathbf{Tr_v} = \{Tr_v^1, Tr_v^2, ..., Tr_v^n\}$ for vehicle $v \in V$. An $i$-th time trace of vehicle $v \in V$ is denoted by $Tr_v^{i} = \{t_{v,i}^1, t_{v,i}^2, ..., t_{v,i}^{|T_v^{i}|}\}$ where each element $t_{v,i}^k \in T$ of $Tr_v^{i}$ is the time stamp of the corresponding GPS point $r_{v,i}^k$ of $R_v^{i}$ (\emph{i.e.,} $|Tr_v^{i}| = |R_v^{i}|$).

There are $N_s$ sensors denoted by $S = \{s_1, s_2, ..., s_{N_s}\}$. The GPS location of a sensor $s \in S$ is denoted by $p_s \in \mathbb{R}^2$. The data transmission cost of a sensor denoted by $\alpha_{cost}$ (dollars) is an amount of money paid to a vehicle for transmitting a data unit. The data transmission cost is determined considering the data subscription plan, \emph{i.e.} it should be smaller than the actual cost for transmitting a data unit to save money each time a data unit is transmitted. Another parameter for a sensor is the communication range denoted by $\alpha_{range}$ in meters. A sensor can only communicate with a vehicle if the geodetic distance between the vehicle and sensor is smaller than $\alpha_{range}$. Lastly, the data generation rate of a sensor is $\alpha_{rate}$ which means that a sensor generates $\alpha_{rate}$ data units per second.

A $N_v \times N_s \times T$ matrix denoted by $\matr{M}^{rx}$ represents the data transmission status between sensors and vehicles during a time period $T$, \emph{i.e.,} indicating whether a sensor has transmitted a data unit to a vehicle or not at any time. More precisely, an element $\matr{M}^{rx}_{i,j,t}$ of matrix $\matr{M}$ is 1 if vehicle $i \in V$ is within the range of sensor $j \in S$ at time $t$ and receives a data unit from sensor $j$ at time $t$. A single data unit represents the amount of data that can be transmitted by a sensor in one time slot which can be easily set up based on the specifications of LPWAN nodes. We also define a $N_v \times N_s \times T$ matrix denoted by $\matr{D}$. Each element $\matr{D}_{i,j,t}$ of this matrix $\matr{D}$ represents the geodetic distance between a vehicle $i \in V$ and a sensor $j \in S$ at time $t$. This matrix $\matr{D}$ can be pre-computed before performing optimization using the communication range $\alpha_{range}$, set of trajectories $\mathbf{R_v}$, and corresponding time traces $\mathbf{Tr_v}$, $\forall v$.

We assume that the operator sets the maximum amount of budget denoted by $c_{max}$ dollars that can be spent to pay for participating vehicles, \emph{i.e.,} if a vehicle relays a data packet received from a sensor, it will receive money. We also assume that each participating vehicle receives at least $c_{min}$ dollars during a time period $T$ to encourage participation. The maximum budget $c_{max}$ for a time period $T$, therefore, should be smaller than the actual data subscription cost to save money. All notations and parameters used throughout this article are summarized in Table II.

\begin{table}[t]
\center
\label{table:notations}
\caption{List of notations}
\begin{tabular}{ l|l }
  \hlinewd{1.5pt}
  \multicolumn{2}{c}{Notations} \\
  \hline
  $T=\{t_1, t_2, ..., t_n\}$ & \pbox{4.5cm}{The discretized time into time slots. The duration of each time slot is 1 second} \\ \hline
  $V=\{v_1, v_2, ..., v_{N_v}\}$ & \pbox{4.5cm}{A set of vehicles wishing to participate in the incentive program} \\ \hline
  $\mathbf{R_v} = \{R_v^1, R_v^2, ..., R_v^{|R_v|}\}$ & \pbox{4.5cm}{A set of trajectories of vehicle $v \in V$} \\\hline
  $R_v^{i} = \{r_{v,i}^1, r_{v,i}^2, ..., r_{v,i}^{|R_v^{i}|}\}$ & \pbox{4.5cm}{An $i$-th trajectory of vehicle $v$ consisting of a sequence of $|R_v^{i}|$ GPS points $r_{v,i}^j \in \mathbb{R}^2$} \\\hline
  $\mathbf{Tr_v} = \{Tr_v^1, ..., Tr_v^{|Tr_v|}\}$ & \pbox{4.5cm}{A set of time traces of vehicle $v \in V$} \\\hline
  $Tr_v^{i} = \{t_{v,i}^1, t_{v,i}^2, ..., t_{v,i}^{|Tr_v^{i}|}\}$ & \pbox{4.5cm}{An $i$-th time trace of vehicle $v$ consisting of a sequence of $|T_v^{i}|$ time stamps corresponding to $R_v^{i}$ (\emph{i.e.,$|T_v^{i}| = |R_v^{i}|$)} } \\\hline
  $S = \{s_1, s_2, ..., s_{N_s}\}$ & \pbox{4.5cm}{A set of deployed LPWAN sensors.} \\\hline
  $p_s \in \mathbb{R}^2$ & \pbox{4.5cm}{The GPS location of sensor $s \in S$} \\\hline
  $\alpha_{cost}$ & \pbox{4.5cm}{A cost for transmitting a data unit to a participating vehicle} \\\hline
  $\alpha_{range}$ & \pbox{4.5cm}{The communication range of a sensor} \\\hline
  $\alpha_{rate}$ & \pbox{4.5cm}{The data generation rate of a sensor} \\\hline
  $c_{min}$ & \pbox{4.5cm}{A minimum compensation being paid to participating vehicles} \\\hline
  $c_{max}$ & \pbox{4.5cm}{The maximum budget} \\\hline
  $\matr{M}^{rx}$ & \pbox{4.5cm}{A $N_v \times N_s \times T$ matrix representing transmissions between vehicles and sensors during a time period $T$} \\\hline
  $\matr{D}$ & \pbox{4.5cm}{A $N_v \times N_s \times T$ matrix representing time-varying geodetic distances between vehicles and sensors during a time period $T$} \\
  \hlinewd{1.5pt}
\end{tabular}
\end{table}


\section{Proposed Approach}
\label{sec:proposed_approach}


\subsection{Overview}
\label{subsec:overview}

\begin{figure}[h]
\centering
\includegraphics[width=.7\columnwidth]{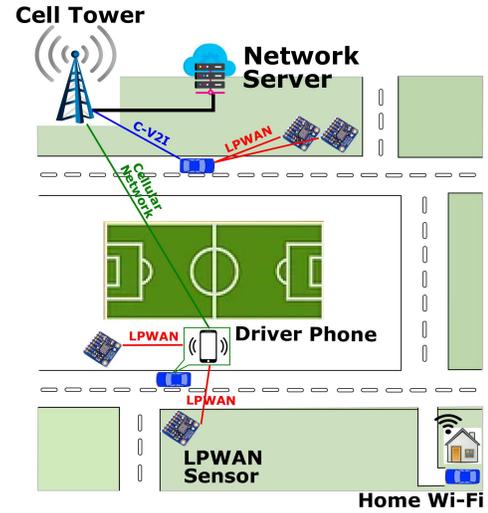}
\caption {An overview of the system architecture for LOC-LPWAN.}
\label{fig:overview}
\end{figure}

An overview of the operation of LOC-LPWAN is presented. As illustrated in Fig.~\ref{fig:overview}, vehicles wishing to participate in the incentive program send a request with their travel schedules to the server via an App or a dedicated website. The server selects vehicles to participate and computes the communication schedule between the selected vehicles and LPWAN sensors to maximize the throughput given an available budget (Section~IV-C). The selected vehicles communicate directly with sensors using CTC, \emph{e.g.,} sensor $\leftrightarrow$ driver phone, sensor $\leftrightarrow$ vehicle's C-V2I module, or sensor $\leftrightarrow$ driver home Wi-Fi, \emph{etc.} based on the communication schedule as illustrated in Fig.~\ref{fig:overview}. Additionally, the server is capable of creating a communication schedule in such a way that no sensor is allowed to dominate the channel by transmitting too many packets compared to other sensors (fairness constraint in Section~IV-D). Furthermore, the operator of LPWAN is provided with an option to ensure that a packet is delivered to the server within a specific time bound (delay constraint in Section~IV-E).

\subsection{Cross Technology Communication}
\label{sec:cross_technology_communication}

The CTC technology is used to enable communication between LPWAN sensors and vehicles. LOC-LPWAN can be integrated with any CTC technology. For example, cellular-V2X (C-V2X) supports two modes of operation using different ISM bands, \emph{i.e.,} 5.9 GHz for vehicle-to-vehicle (V2V) and 2 GHz for V2I~\cite{yusuf2020experimental}\cite{liu2020joint}. The ISM band for the narrowband IoT (NB-IoT) which is also called as LTE Cat NB1 for LPWAN service is between 450 MHz and 2200 MHz~\cite{nbiot-band}. An interesting observation is that the two heterogeneous communication technologies share the ISM band allowing for the opportunity for development of the CTC technology to enable seamless communication between LPWAN sensors and vehicles.

\begin{figure}[h]
\centering
\includegraphics[width=.9\columnwidth]{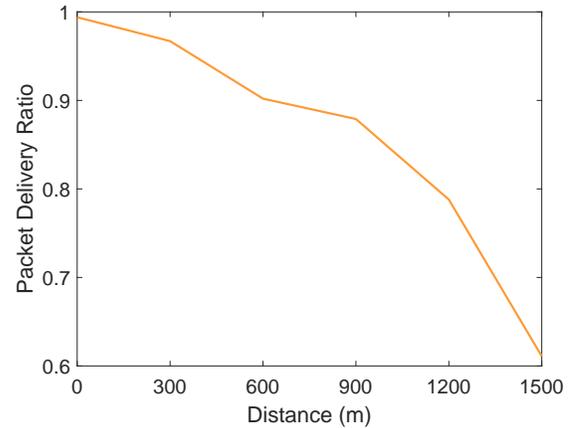}
\caption {The packet delivery ratio (PDR) of the proof-of-concept CTC system based on a driver smartphone.}
\label{fig:CTC_Proofofconcept}
\end{figure}

A driver phone can also be used to enable CTC between LPWAN nodes and vehicles. We demonstrate the feasibility of LOC-LPWAN integrated with a CTC technology by implementing a proof-of-concept CTC technology based on a driver smartphone. More specifically, a LPWAN node is implemented with an Arduino board equipped with a NBIoT module, u-Blox SARA-R410M-02B~\cite{ublox}. A driver phone for CTC is implemented with Motorola Moto G5 running Android version 9. A Hologram IoT simcard is integrated to use the NB-IoT channel. To evaluate the feasibility, we measure the packet delivery ratio (PDR) by varying distance between the two devices. Results are depicted in Fig.~\ref{fig:CTC_Proofofconcept} demonstrating that the two devices reliably exchange packets with over 90\% PDR within the distance of 600m which is long enough for communication between LPWAN sensors and vehicles.

An interesting aspect of LOC-LPWAN is that the communication schedule can be updated frequently to take into account the current traffic conditions. Especially, when participating vehicles fail to show up, the schedule can be quickly updated to select a new vehicle to relay packets. In particular, we explain in more detail in Section~\ref{sec:execution_time} that such a frequent schedule update is feasible by demonstrating the small execution time for obtaining the updated solution.

\subsection{Communication Scheduling}
\label{subsec:scheduling}

An optimization problem is formulated to find the optimal communication schedule between sensors and vehicles such that the total amount of data transmitted by sensors is maximized given a set of parameters $\{\alpha_{cost}, \alpha_{range}, \alpha_{rate}, c_{min}, c_{max}\}$. Solving this optimization problem allows the operator to obtain a set of participating vehicles and an amount of money provided to each participating vehicle. The set of participating vehicles is represented as a vector $\bar{V}$ where $\bar{V}_i = 1$ if $\sum\limits_{j \in S} \sum\limits_{t \in T} \matr{M}^{rx}_{i,j,t} \cdot \alpha_{cost} > c_{min}, i \in V$, and otherwise $\bar{V}_i = 0$. This means that a vehicle is considered as a participating vehicle if it receives data from sensors and the amount of compensation is greater than $c_{min}$.

The proposed communication scheduling problem is defined as following.


\begin{definition}
\textbf{Communication Scheduling for Participating Vehicles (CSPV) Problem:} Find the optimal communication schedule between sensors in $S$ and vehicles in $V$ such that the total amount of data transmitted, $\sum\limits_{i \in V} \sum\limits_{j \in S} \sum\limits_{t \in T} \matr{M}^{rx}_{i,j,t}$, is maximized given the trajectories $\mathbf{R_v}$ and time traces $\mathbf{Tr_v}$ of vehicles and the parameter set $\{\alpha_{cost}, \alpha_{range}, \alpha_{rate}, c_{min}, c_{max}\}$. This solution is a set of participating vehicles $\bar{V}$ and an amount of compensation provided to each vehicle, \emph{i.e.,} $\sum\limits_{j \in S} \sum\limits_{t \in T} \matr{M}^{rx}_{i,j,t} \cdot \alpha_{cost}, \forall i \in \bar{V}$.
\end{definition}

The CSPV problem can be formulated as an 0-1 integer linear program as follows.

\begin{displaymath}
\begin{aligned}
\argmaxA_{\bar{V}} \quad & \sum\limits_{i \in V} \sum\limits_{j \in S} \sum\limits_{t \in T} \matr{M}^{rx}_{i,j,t}\\
\textrm{s.t.} \quad & \sum\limits_{i \in V} \sum\limits_{j \in S} \sum\limits_{t \in T} \matr{M}^{rx}_{i,j,t} \cdot \alpha_{cost} \leq c_{max} & (1)\\
& \sum\limits_{j \in S} \sum\limits_{t \in T} \matr{M}^{rx}_{i,j,t} \cdot \alpha_{cost} > c_{min}, \forall i \in \bar{V} & (2)\\
& 0 \leq \sum\limits_{i \in V} \matr{M}^{rx}_{i,j,t} \leq 1, \forall j, t & (3)\\
& \sum\limits_{i \in V} \sum\limits_{t' < t} \matr{M}^{rx}_{i,j,t'} \le t \cdot \alpha_{rate}, \forall j, t& (4)\\
& \matr{D}_{i,j,t} \cdot \matr{M}^{rx}_{i,j,t} \leq \alpha_{range}, \forall i,j,t & (5)\\
& 0 \le \bar{V}_i \le 1, \forall i & (6)\\
& \bar{V}_i = min(\sum\limits_{j \in S} \sum\limits_{t \in T} \matr{M}^{rx}_{i,j,t}, 1), \forall i& (7)\\
& 0 \le \matr{M}^{rx}_{i,j,t} \le 1, \forall i,j,t& (8)\\
\end{aligned}
\end{displaymath}

\noindent where Constraints (1) and (2) are related to the budget. Specifically, Constraint (1) ensures that the total amount of compensation provided to all participating vehicles in $\bar{V}$ should be smaller than or equal to the maximum budget $c_{max}$; Constraint (2) specifies the minimum amount of compensation, \emph{i.e.,} at least $c_{min}$ should be given to each participating vehicle; This constraint is introduced to encourage participation of vehicles. Constraint (3) ensures that a sensor can communicate with only one vehicle at a time (\emph{i.e.,} unicast) to prevent vehicles from transmitting the same packet to save the cost. Constraint (4) enforces that a sensor can only transmit data if it has data to send in its buffer. Whether or not a sensor has data in its buffer is determined based on the data generation rate $\alpha_{rate}$. Constraint (5) dictates that a sensor can only communicate with a vehicle within its communication range $\alpha_{range}$. Note that $\matr{M}^{rx}_{i,j,t}  = 1$ if communication occurs between vehicle $i \in V$ and sensor $j \in S$ at time $t$; thus $\matr{D}_{i,j,t} \cdot \matr{M}^{rx}_{i,j,t}$ means the geodetic distance between vehicle $i$ and sensor $j$ at time $t$ when communication happens between them. A more realistic communication range can be easily incorporated into our problem by replacing $\alpha_{range}$ with a function that represents a particular communication range model. Constraints (6) and (7) are used to define the set of participating vehicles. Constraint (8) is the boundary conditions of variables. Once $\bar{V}$ is identified by solving the CSPV problem, the amount of compensation provided to each participating vehicle can be calculated based on the number of data units forwarded by the vehicle.

One of the interesting features of LOC-LPWAN is that it can take into account the dynamically changing trajectories of vehicles (\emph{i.e.,} $\mathbf{R_v}$ and $\mathbf{Tr_v}$) in response to the current traffic conditions. More specifically, LOC-LPWAN allows to update the participating vehicles and recalculate the amount of compensation for each vehicle according to dynamically changing traffic conditions.
Furthermore, a history of trajectories for the same trip over multiple days can be used to improve the trajectory information of a vehicle. More precisely, a ``mean'' trajectory constructed with the average arrival times at a set of predefined GPS locations of a trip using the history of trajectories can be used instead of adopting a certain trajectory for a specific date to mitigate the impact of day-to-day traffic dynamics.

\subsection{Ensuring Fairness Among Sensors}
\label{subsec:fairness}

While a solution for the CSPV problem determines the participating vehicles and the amount of compensation for each of those vehicles, it does not guarantee fairness among sensors in terms of the amount of data that the sensors transmit. For example, we may end up with a solution where some sensors transmit substantially more data units than other sensors. To address this fairness issue, we extend the CSPV problem and define a new problem as follows.

\begin{definition}
\textbf{Fair Communication Scheduling for Participating Vehicles (F-CSPV) Problem:} To guarantee fairness among sensors, solve the CSPV problem such that the difference between the largest number of data units transmitted by a sensor, $\mbox{max}_{j \in S}\sum\limits_{i \in V}\sum\limits_{t \in T}\matr{M}^{rx}_{i,j,t}$, and the smallest number of data units transmitted by a sensor, $\mbox{min}_{j \in S}\sum\limits_{i \in V}\sum\limits_{t \in T}\matr{M}^{rx}_{i,j,t}$, is minimized.
\end{definition}

We formulate the F-CSPV problem by modifying the objective function of the CSPV problem and introducing a new parameter $\digamma (0 \le \digamma \le 1)$ that is used to control the degree of fairness as follows.

\begin{displaymath}
\argmaxA_{\bar{V}} \quad \Bigg[ \digamma \cdot \frac{\sum\limits_{i \in V}\sum\limits_{j \in S}\sum\limits_{t \in T}\matr{M}^{rx}_{i,j,t}}{|S||V||T|}
\end{displaymath}

\begin{displaymath}
- (1 - \digamma) \cdot \frac{\mbox{max}_{j \in S}\sum\limits_{i \in V}\sum\limits_{t \in T}\matr{M}^{rx}_{i,j,t} - \mbox{min}_{j \in S}\sum\limits_{i \in V}\sum\limits_{t \in T}\matr{M}^{rx}_{i,j,t}}{|V||T|}\Bigg]
\end{displaymath}

\begin{displaymath}
\textrm{s.t.} \quad  \mbox{All constraints} \quad (1)-(8)
\end{displaymath}

\noindent Note that the objective function is designed to strike a balance between throughput and fairness by introducing a new parameter $\digamma$. More specifically, the first term of the objective function (\emph{i.e.,} $\digamma \cdot \frac{\sum\limits_{i \in V}\sum\limits_{j \in S}\sum\limits_{t \in T}\matr{M}^{rx}_{i,j,t}}{|S||V||T|}$) represents the throughput, and the second term (\emph{i.e.,} $(1 - \digamma) \cdot \frac{\mbox{max}_{j \in S}\sum\limits_{i \in V}\sum\limits_{t \in T}\matr{M}^{rx}_{i,j,t} - \mbox{min}_{j \in S}\sum\limits_{i \in V}\sum\limits_{t \in T}\matr{M}^{rx}_{i,j,t}}{|V||T|}$) means the fairness. These terms are normalized to fit in the range between 0 and 1. We then linearize the objective function by introducing two new variables $z_{max}$ and $z_{min}$ and adding new constraints to remove the max and min functions of the objective function as follows.

\begin{displaymath}
\begin{aligned}
\argmaxA_{\bar{V}} \quad & \Bigg[ \digamma \cdot \frac{\sum\limits_{i \in V}\sum\limits_{j \in S}\sum\limits_{t \in T}\matr{M}^{rx}_{i,j,t}}{|S||V||T|} - (1 - \digamma) \cdot \frac{z_{max} - z_{min}}{|V||T|}\Bigg]\\
\textrm{s.t.} \quad & \mbox{All constraints} \quad\quad\quad\quad\quad\quad\quad\quad\quad (1)-(8)&\\
& z_{max} \ge \sum\limits_{i \in V}\sum\limits_{t \in T}\matr{M}^{rx}_{i,j,t}, \forall j \in S \quad\quad\quad\quad (9) &\\
& z_{min} \le \sum\limits_{i \in V}\sum\limits_{t \in T}\matr{M}^{rx}_{i,j,t}, \forall j \in S \quad\quad\quad\quad (10)&\\
\end{aligned}
\end{displaymath}

\subsection{Delay-Bound Communication Scheduling}
\label{subsec:real_time}

We note that some applications are not delay tolerant mandating sensor data to be transmitted within a certain time bound. To impose such a delay bound, we define a new problem namely the Delay-bound F-CSPV (DF-CSPV) problem as follows.

\begin{definition}
\textbf{Delay-Bound and Fair Communication Scheduling for Participating Vehicles (DF-CSPV) Problem:} Solve the F-CSPV problem such that any data unit should be transmitted within a certain delay bound $\delta$.
\end{definition}

The DF-CSPV problem is formulated based on the F-CSPV problem by adding a new constraint defined with parameter $\delta$ as follows.

\begin{displaymath}
\begin{aligned}
\argmaxA_{\bar{V}} \quad & \Bigg[ \digamma \cdot \frac{\sum\limits_{i \in V}\sum\limits_{j \in S}\sum\limits_{t \in T}\matr{M}^{rx}_{i,j,t}}{|S||V||T|} - (1 - \digamma) \cdot \frac{z_{max} - z_{min}}{|V||T|}\Bigg]\\
\textrm{s.t.} \quad & \mbox{All constraints} \quad\quad\quad\quad\quad\quad\quad\quad\quad (1)-(10)&\\
& \matr{M}^{rx}_{i,j,t} \cdot t - \frac{\sum\limits_{i' \in V} \sum\limits_{t' \le t} \matr{M}^{rx}_{i',j,t'}}{\alpha_{rate}} < \delta, \forall i,j,t \quad (11)&\\
\end{aligned}
\end{displaymath}

\noindent Here, since $\matr{M}^{rx}_{i,j,t}$ is 1 if sensor $j \in S$ transmitted a data unit to vehicle $i \in V$ at time $t$, $\matr{M}^{rx}_{i,j,t} \cdot t$ is the time when the data unit was transmitted, acquiring the `transmission time'. We then subtract the `data generation time' (\emph{i.e.,} the time when the data unit was generated) from the transmission time to obtain the delay. Specifically, $\sum\limits_{i' \in V} \sum\limits_{t' < t} \matr{M}^{rx}_{i',j,t'}$ is the total number of data units that have been transmitted by sensor $j$ until time $t$. If we divide this by the data generation rate $\alpha_{rate}$, we get the time when the latest data unit was generated, acquiring the `data generation time'. Consequently, we can calculate the delay as the transmission time minus the data generation time. This delay should be smaller than $\delta$ to satisfy the delay-bound constraint. For example, assume that a sensor transmits a data unit to a car at 5 second. So, the transmission time is 5 (\emph{i.e.,} $\matr{M}^{rx}_{i,j,t} \cdot t = 5$). Also assume that the data generation rate $\alpha_{rate}$ is 2, which means that 2 data units are generated per second and that so far, the sensor has transmitted 3 data units (\emph{i.e.,} $\sum\limits_{i' \in V} \sum\limits_{t' < t} \matr{M}^{rx}_{i',j,t'} = 3$). The generation time for the latest data unit that is just transmitted is $3/2 = 1.5$. Thus, the time difference between the transmission time and the data generation time is = 5 - 1.5 = 3.5 second which is the delay.



\section{Results}
\label{sec:evaluation}

A simulation study is performed to evaluate the performance of LOC-LPWAN. We first describe the dataset used for this simulation study, and the algorithms compared with LOC-LPWAN are explained. We then present an analysis on the throughput of LOC-LPWAN. After we determine the optimal value for parameter $\digamma$, we analyze the degree of fairness using the parameter. We also validate whether LOC-LPWAN allows vehicles to complete packet transmission within a specific delay bound. Finally, the computation time for LOC-LPWAN is measured to demonstrate the feasibility of updating solutions quickly in response to dynamically changing traffic conditions.

\subsection{Simulation Setup}
\label{sec:experimental_setup}

We use a PC equipped with Intel Core  i7-9750H CPU @ 2.60 GHz with 16 GB of RAM running the Windows 10 64-bit operating system. To implement, run, and obtain solutions for the CSPV, F-CSPV, and DF-CSPV problems, we use the MATLAB optimization toolbox~\cite{matlab}. We adopt real-world vehicle trajectories, \emph{i.e.,} the T-Drive trajectory dataset~\cite{yuan2011driving}\cite{yuan2010t} that contains one-week trajectories of taxis with the total length of 9 million kilometers. From this dataset, we extract day-long GPS trajectories of 40 taxis within 6 districts in Beijing, China, considering that LOC-LPWAN can be utilized most effectively for cost savings in a region-by-region basis for the purpose of quick solution updates, rather than covering the whole city. We then deploy 10 sensors randomly in the target region. We repeat deployment of these sensors 10 times, thereby creating 10 random scenarios for this analysis. The parameters $\alpha_{cost}, \alpha_{range}, \alpha_{rate}, c_{min},$ and $ c_{max}$ are set to \$1/MB, 2,000 m, 1 KB/sec, \$2, and \$1,000, respectively. Given the parameters, we measure the throughput, degree of fairness, and delay for packet transmission.


Since LOC-LPWAN is the first approach designed to compute the optimal communication schedule between sensors and vehicles based on the CTC technology, we create a greedy algorithm to compare the performance with LOC-LPWAN. Specifically, we create a greedy algorithm that allows sensors to transmit data units whenever there is a vehicle within its communication range. The pseudocode of the greedy algorithm is shown in Algorithm~\ref{algorithm1}. In this algorithm, each sensor keeps generating sensor data according to its data generation rate $\alpha_{rate}$ (Line 6), where $Data_s$ is the number of data units that have been generated by sensor $s$. If sensor $s$ has data units to transmit and there is a vehicle $v$ within the communication range $\alpha_{range}$ (Lines 7-9), sensor $s$ transmits a data unit to vehicle $v$. The algorithm terminates when the available budget $c_{max}$ is exhausted (Line 16-17). Note that $\mathbf{R_v}$ and $\mathbf{Tr_v}$ are used to determine whether a vehicle is within the range of a sensor at a certain time or not. Lastly, the greedy algorithm excludes the vehicles from the obtained list of participating vehicles that have not received enough compensation $c_{min}$ to make it consistent with LOC-LPWAN.

\begin{algorithm}
  \label{algorithm1}
  \caption{The Greedy Algorithm}
  \textbf{Input:} $T, \mathbf{R_v}, \mathbf{Tr_v}, V, S, \alpha_{cost}, \alpha_{range}, \alpha_{rate}, c_{min}, c_{max}$ \;
  \textbf{Output:} $\matr{M}^{rx}$ \;
  \For {each time $t \in T$} {
    \For {each sensor $s \in S$} {
        // Generate sensor data according to the data generation rate $\alpha_{rate}$. \;
        $Data_{s} \leftarrow Data_{s} + \alpha_{rate}$ \;
        \If {$Data_{s} \ge 1$} {
            \For {each vehicle $v \in V$} {
                \If {$v$ is within the range $\alpha_{range}$ of $s$} {
                    // If vehicle $v$ has not received data units from any sensor yet at time $t$, then allow sensor $s$ to transmit a data unit to vehicle $v$. \;
                    \If {$\matr{M}^{rx}_{v,j,t} \neq 1, \forall j$} {
                        $\matr{M}^{rx}_{v,s,t} \leftarrow 1$ \;
                        $Data_s \leftarrow Data_{s} - 1$ \;
                        break
                    }
                }
            }
        }
        // Terminate if the budget is used up. \;
        \If {$\alpha_{cost} \cdot \sum\limits_{i \in V} \sum\limits_{j \in S} \sum\limits_{t \in T} \matr{M}^{rx}_{i,j,t} > c_{max}$} {
            return \;
        }
    }
  }
\end{algorithm}

Since the greedy algorithm excludes the vehicles receiving the compensation smaller than $c_{min}$, if we accumulate those canceled payments, we can create a new budget. This new budget can be used to hire more vehicles by rerunning the greedy algorithm, potentially improving the throughput, fairness, and delay. For example, say that three vehicles $v_1$, $v_2$, $v_3$ are excluded after running the greedy algorithm because the amount of compensation for these vehicles denoted by $c_1$, $c_2$, and $c_3$, respectively, are all smaller than $c_{min}$. We can add up these payments and create a new budget $c_{new}$, \emph{i.e.,} $c_{new} = c_1+c_2+c_3$, and we can use this budget to find more participating vehicles by rerunning the greedy algorithm. This process can be repeated until $c_{new}$ becomes smaller than $c_{min}$, continually improving the performance. In this article, we denote the original greedy algorithm by `Greedy' and this method of repeatedly running the greedy algorithm by 'Greedy-N'. Besides comparing the performance of LOC-LPWAN with these greedy algorithms, we also compare the performance with the baseline algorithm where no vehicle is used to relay the packets.

\subsection{Throughput}
\label{sec:more}

\begin{figure}[h]
\centering
\includegraphics[width=.9\columnwidth]{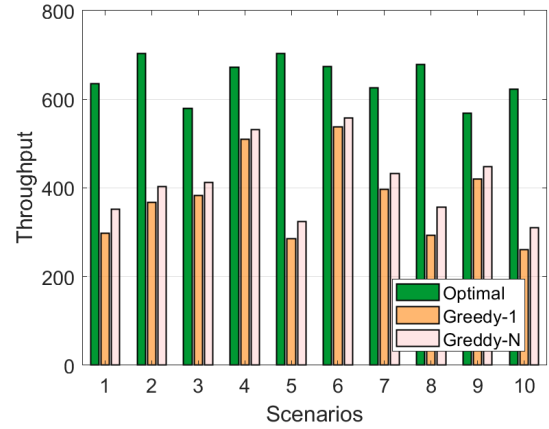}
\caption {The total number of data units transferred for LOC-LPWAN, greedy, and greedy-N algorithms.}
\label{fig:selected_vehicles_data}
\end{figure}

LOC-LPWAN is designed to maximize the throughput represented by the number of data units transmitted during a specific time period given the limited budget. In this section, we measure the throughput in comparison with the greedy, greedy-N, and baseline algorithms in 10 random scenarios. The results are displayed in Fig.~\ref{fig:selected_vehicles_data}. As shown, LOC-LPWAN has remarkably higher throughput in all scenarios than the greedy and greedy-N algorithms by up to 146.7\% and 116\%, respectively.

\begin{figure}[h]
\centering
\includegraphics[width=.9\columnwidth]{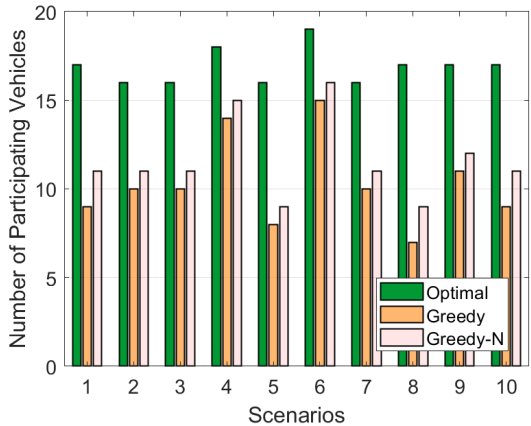}
\caption {The number of participating vehicles for LOC-LPWAN, greedy, and greedy-N algorithms in 10 random scenarios.}
\label{fig:selected_vehicles_num_participants}
\end{figure}

We then measure the number of participating vehicles for LOC-LPWAN and compare with the greedy and greedy-N algorithms. The results are depicted in Fig.~\ref{fig:selected_vehicles_num_participants}. Although more vehicles are selected by LOC-LPWAN in all scenarios than that of both the greedy and greedy-N algorithms, the number of selected participating vehicles vary in different scenarios. The reason is because the proposed optimization solution is focused on maximizing the total amount of data transmitted without considering the number of participating vehicles. In other words, it cannot differentiate selecting a relatively small number of participating vehicles with a large amount of data transmissions from selecting a large number of participating vehicles with a small amount of data transmissions. However, It is worthy to note that if the operator needs to impose a constraint on the maximum number of data units transmitted to each participating vehicle, it is easy to add such a constraint, \emph{i.e.,} $\sum\limits_{j \in S} \sum\limits_{t \in T} \matr{M}^{rx}_{i,j,t} < N, \forall i \in V$.

\begin{figure}[h]
\centering
\includegraphics[width=.9\columnwidth]{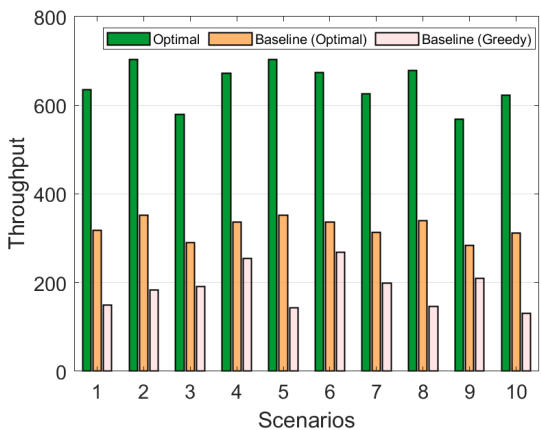}
\caption {The throughput of LOC-LPWAN compared with the baseline approach where no vehicle is used for relaying packets.}
\label{fig:selected_vehicles_data_base}
\end{figure}

Now we compare the throughput of LOC-LPWAN with the baseline algorithm where no vehicle is used to relay packets. More precisely, we measure the operation cost $c_{op}^{opt}$ (\emph{i.e.,} compensation given to participating vehicles) for LOC-LPWAN. We then measure the number of data units transmitted by the baseline approach assuming that the same amount of money $c_{op}^{opt}$ is available based on the cost model of \$1/1 MB. The results demonstrate that LOC-LPWAN achieves 100.2\% higher throughput on average compared with the baseline approach as depicted in Fig.~\ref{fig:selected_vehicles_data_base}. The results indicate that it is significantly more economical to use vehicles to relay packets than relying on the fixed LPWAN network.

We conduct the same experiment for the greedy algorithm, \emph{i.e.,} comparing the cost between the greedy algorithm and the baseline approach. We observe that the throughput achieved by the greedy algorithm is significantly higher by 100.5\% compared with the baseline approach with the same amount of budget, indicating that the greedy algorithm also achieves significant cost reduction. Furthermore, we note that such cost saving can be increased by adjusting the minimum amount of compensation paid to the participating vehicles. Overall, with the same amount of budget, both the LOC-LPWAN and the greedy algorithm achieve nearly 2X higher throughput, indicating the cost reduction of nearly 50\%.

\subsection{Fairness}
\label{sec:param_fairness}

\begin{figure}[h]
\centering
\includegraphics[width=.9\columnwidth]{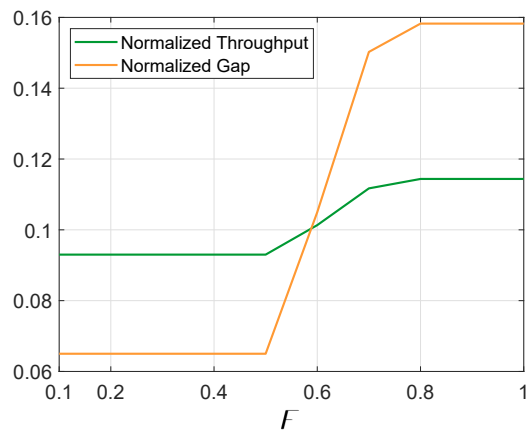}
\caption {The effect of $\digamma$ on the throughput and delay.}
\label{fig:effect_of_alpha}
\end{figure}

We evaluate the performance of LOC-LPWAN in terms of the degree of fairness for data transmission among sensors. We first determine the optimal value of parameter $\digamma$ for each scenario to use it for measuring the degree of fairness. Specifically, we aim to find the value of $\digamma$ that maximizes the normalized throughput (\emph{i.e.,} $\frac{\sum\limits_{i \in V}\sum\limits_{j \in S}\sum\limits_{t \in T}\matr{M}^{rx}_{i,j,t}}{|S||V||T|}$) and minimizes the normalized gap between the largest and smallest amount of data units transmitted by sensors (\emph{i.e.,} $\frac{\mbox{max}_{j \in S}\sum\limits_{i \in V}\sum\limits_{t \in T}\matr{M}^{rx}_{i,j,t} - \mbox{min}_{j \in S}\sum\limits_{i \in V}\sum\limits_{t \in T}\matr{M}^{rx}_{i,j,t}}{|V||T|}$). Recall that the smaller gap means higher fairness. Fig.~\ref{fig:effect_of_alpha} depicts the results. As the value of $\digamma$ increases, the throughput is enhanced since higher $\digamma$ means stronger weight on the throughput; however, at the cost of increased throughput, the normalized gap increases, meaning that the degree of fairness degrades. The results suggest that there is a tradeoff between the throughput and fairness and demonstrate that one can easily choose the value of $\digamma$ that maximizes the objective function of the F-CSPV problem.

\begin{figure}[h]
\centering
\includegraphics[width=.9\columnwidth]{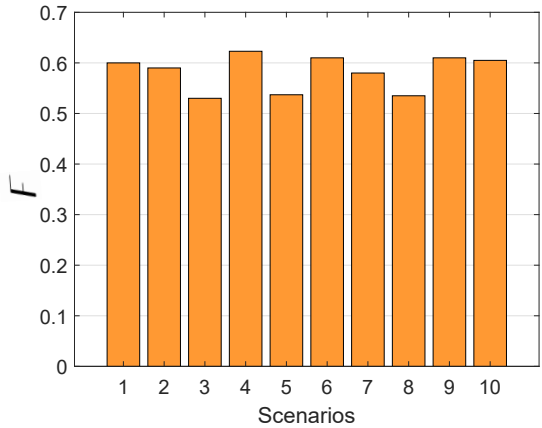}
\caption {Selected $\digamma$ for each scenario.}
\label{fig:selected_alpha}
\end{figure}

Fig.~\ref{fig:selected_alpha} displays the selected value of $\digamma$ for each scenario. It is observed that in most scenarios, the value of $\digamma$ is selected between 0.5 and 0.6 indicating that the normalization of the throughput and delay in the objective function of the F-CSPV problem is adequately defined to balance between the throughput and fairness.

\begin{figure}[h]
\centering
\includegraphics[width=.9\columnwidth]{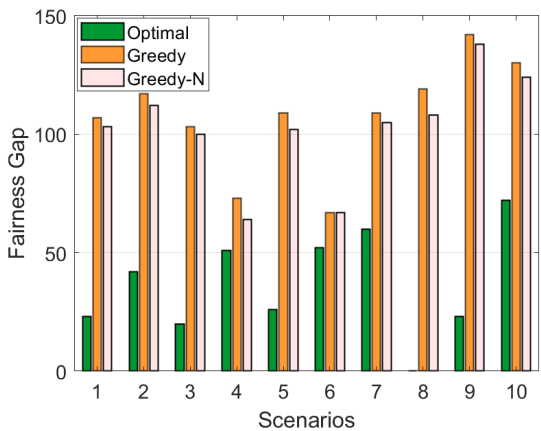}
\caption {The gap between the largest and smallest data units being transmitted by sensors for all scenarios.}
\label{fig:fairness_index}
\end{figure}

Now using the selected value of $\digamma$, we measure the fairness of LOC-LPWAN and compare that with the greedy and greedy-N algorithms. Specifically, we measure the gap between the largest and smallest amount of data transmitted by sensors (\emph{i.e.,} $\mbox{max}_{j \in S}\sum\limits_{i \in V}\sum\limits_{t \in T}\matr{M}^{rx}_{i,j,t} - \mbox{min}_{j \in S}\sum\limits_{i \in V}\sum\limits_{t \in T}\matr{M}^{rx}_{i,j,t}$). The results are depicted in Fig.~\ref{fig:fairness_index}, clearly demonstrating that LOC-LPWAN achieves significantly higher fairness compared with both the greedy and greedy-N algorithms. Specifically, LOC-LPWAN decreases the gap by 65.7\% and 63.9\% on average compared with the greedy and greedy-N algorithms, respectively, and in some case, by almost 100\%.

\subsection{Delay}
\label{sec:test_for_realtime}

\begin{figure}[h]
\centering
\includegraphics[width=.9\columnwidth]{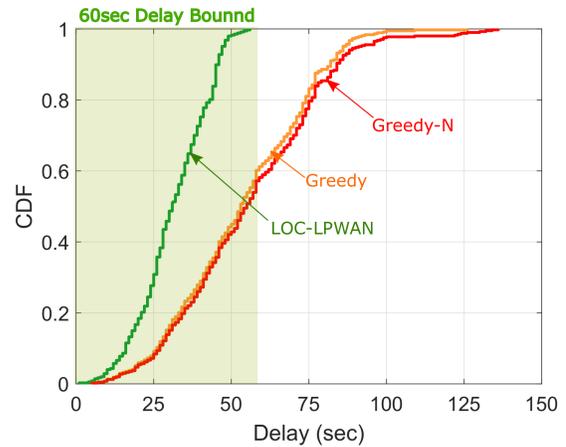}
\caption {Cumulative distribution function (CDF) graph of the delay measured for LOC-LPWAN and the greedy algorithm.}
\label{fig:delay_cdf}
\end{figure}

We evaluate the performance of LOC-LPWAN focusing on the delay-bound scheduling in comparison with the greedy algorithm. We measure the delay between the data generation time and data transmission time for each data unit transmitted. The delay bound $\delta$ is set to 60 seconds, \emph{i.e.,} the communication schedule is determined in such a way that all data units should be sent out from sensors within 60 seconds from the data generation time. Fig~\ref{fig:delay_cdf} depicts the cumulative distribution function (CDF) graph of the delays of all data units transmitted for LOC-LPWAN, greedy, and greedy-N algorithms. The results demonstrate that LOC-LPWAN allows sensors complete data transmission within the specific delay bound. More specifically, compared with the greedy algorithm, LOC-LPWAN reduces the delay by 28.8\% on average. However, it should also be noted that a tight delay constraint reduces the throughput, \emph{i.e.,} with the 60 second of delay constraint, LOC-LPWAN ended up transmitting 15\% smaller number of data units. Another interesting observation is that the delay of Greedy-N is slightly higher by 4.5\% than the greedy algorithm possibly due to the increased throughput and fairness.

\begin{figure}[h]
\centering
\includegraphics[width=.85\columnwidth]{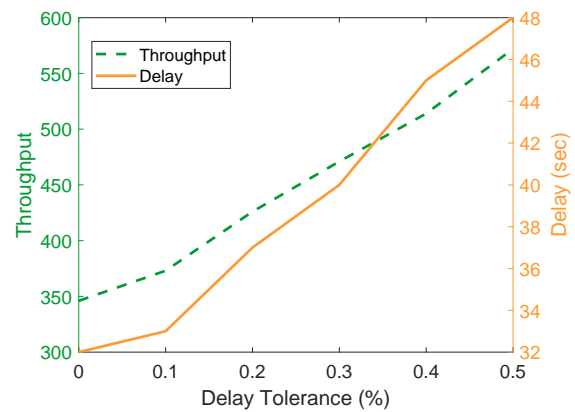}
\caption {Effect of delay tolerance (\%) on throughput and delay.}
\label{fig:ErrorTolerance}
\end{figure}

It is worth to mention that the uncertainty of vehicle movement may affect the performance of LOC-LPWAN. Although LOC-LPWAN guarantees delay-bound data transmission, the throughput can be degraded to meet the delay constraint because sensors will not transmit packets if the delay is expected to be greater than the threshold. A possible idea to address this uncertainty issue is to incorporate delay tolerance, \emph{i.e.,} instead of using a fixed delay bound $\delta$, a new parameter $\alpha_{delay}$ to impose delay tolerance is introduced. More specifically, the constraint (11) can be rewritten as $\matr{M}^{rx}_{i,j,t} \cdot t - \frac{\sum\limits_{i' \in V} \sum\limits_{t' \le t} \matr{M}^{rx}_{i',j,t'}}{\alpha_{rate}} < \delta + \delta\alpha_{delay}, \forall i,j,t$. To understand better the effect of the new parameter, we measure the throughput and delay of LOC-LPWAN by varying $\alpha_{delay}$. The results are shown in Fig.~\ref{fig:ErrorTolerance}. It is observed that as the delay tolerance increases, the throughput increases. The reason is because sensors can transmit more packets because the delay constraint is now relaxed. However, we also observe that the delay increases as the delay tolerance increases as we allow for higher delay tolerance.

\subsection{Penetration Rate}
\label{subsec:penetration_rate}

\begin{figure}[h]
\centering
\begin{minipage}{.48\columnwidth}
\centering
\includegraphics[width=\textwidth]{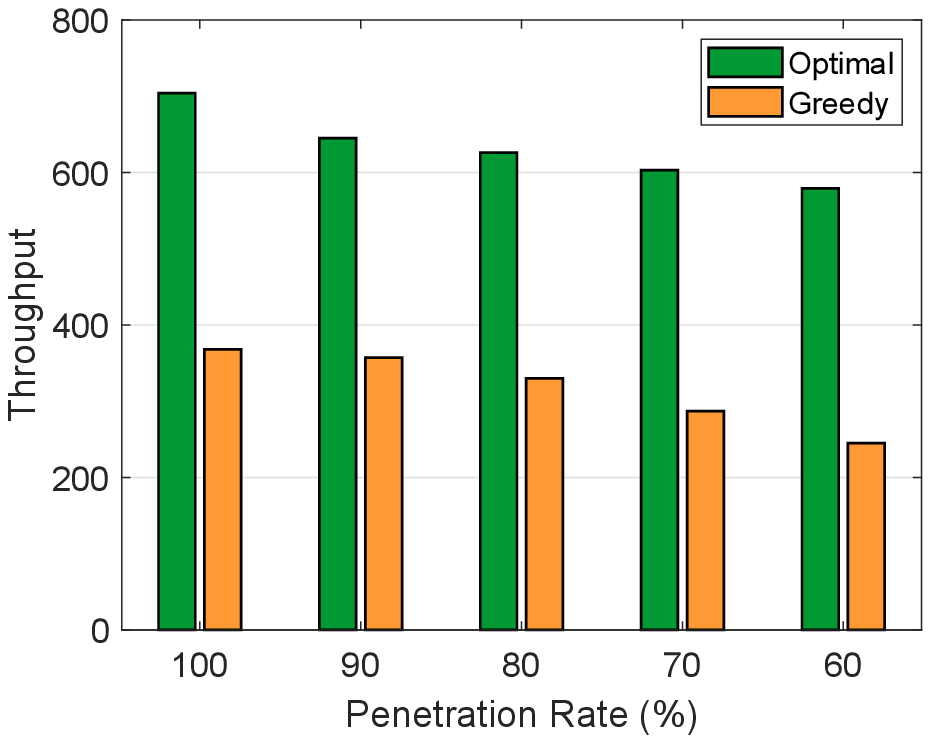}
\vspace{-15pt}
\caption {Effect of penetration rate on throughput.}
\label{fig:PenetrationVsThroughput}
\end{minipage}%
\hspace*{1mm}
\begin{minipage}{.48\columnwidth}
\centering
\includegraphics[width=\textwidth]{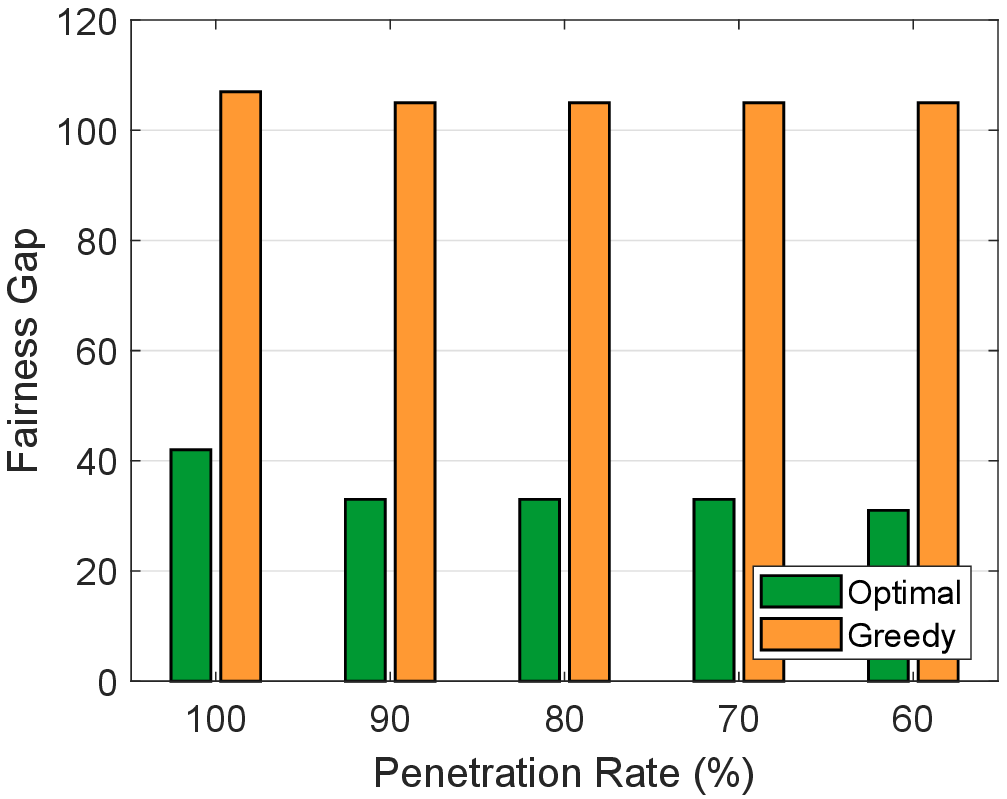}
\vspace{-15pt}
\caption {Effect of penetration rate on fairness.}
\label{fig:PenetrationVsFairness}
\end{minipage}
\end{figure}

Vehicles scheduled to relay packets may not appear at the expected time. The penetration rate is defined as the percentage of vehicles that show up at the scheduled time. In this section, we evaluate the performance of LOC-LPWAN by varying the penetration rate. First, we measure the throughput by varying the penetration rate. In this experiment, the scenario 2 is used with the default parameters. Fig.~\ref{fig:PenetrationVsThroughput} depicts the results demonstrating that the throughput decreases as the penetration rate decreases for both the optimal and greedy algorithms. The reason is that as the penetration rate decreases, some vehicles do not appear resulting in undelivered packets, thereby reducing the overall throughput. We also analyze the effect of the penetration rate on the fairness of data transmission. Fig.~\ref{fig:PenetrationVsFairness} shows the results. An interesting observation is that the effect of penetration rate on the degree of fairness is not decisive. While it is generally anticipated that the degree of fairness does not change much as the penetration rate decreases, we observe that the degree of fairness can increase as the penetration rate decreases. The reason is because as the penetration rate decreases, it is possible that a vehicle scheduled to deliver the largest/smallest number of packets does not appear, thereby making the difference between the minimum and maximum number of packet transmissions smaller, thereby increasing the degree of fairness (smaller fairness gap).

\begin{figure}[h]
\centering
\begin{minipage}{.48\columnwidth}
\centering
\includegraphics[width=\textwidth]{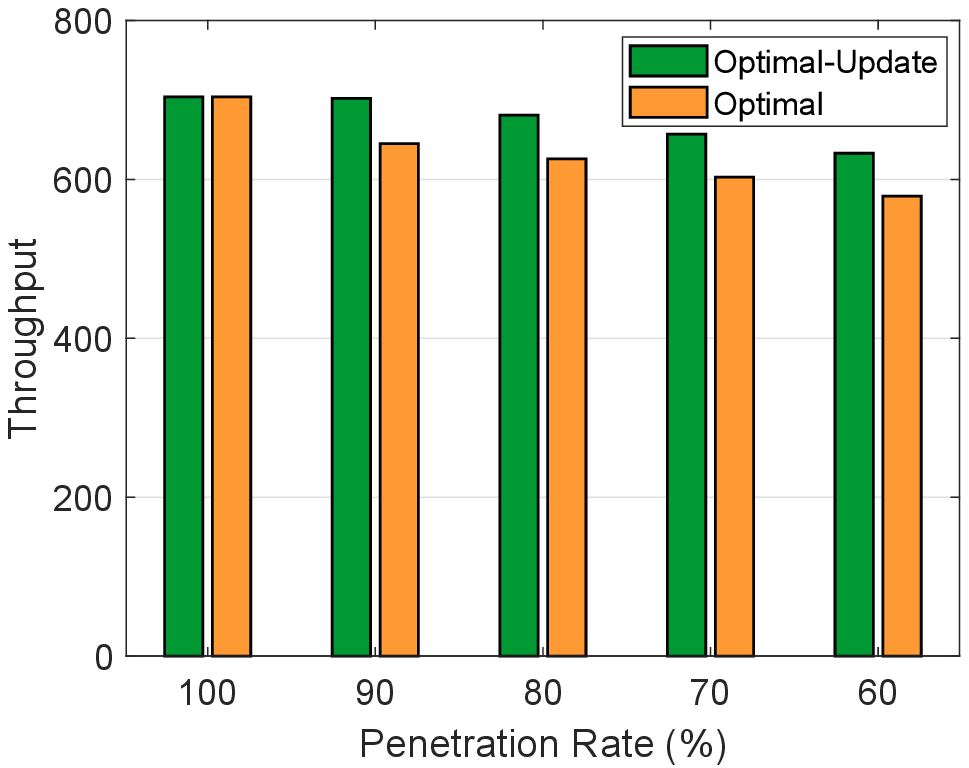}
\vspace{-15pt}
\caption {Effect of penetration rate on throughput after recomputing the communication schedule.}
\label{fig:PenetrationVsThroughputAdvanced}
\end{minipage}%
\hspace*{1mm}
\begin{minipage}{.48\columnwidth}
\centering
\includegraphics[width=\textwidth]{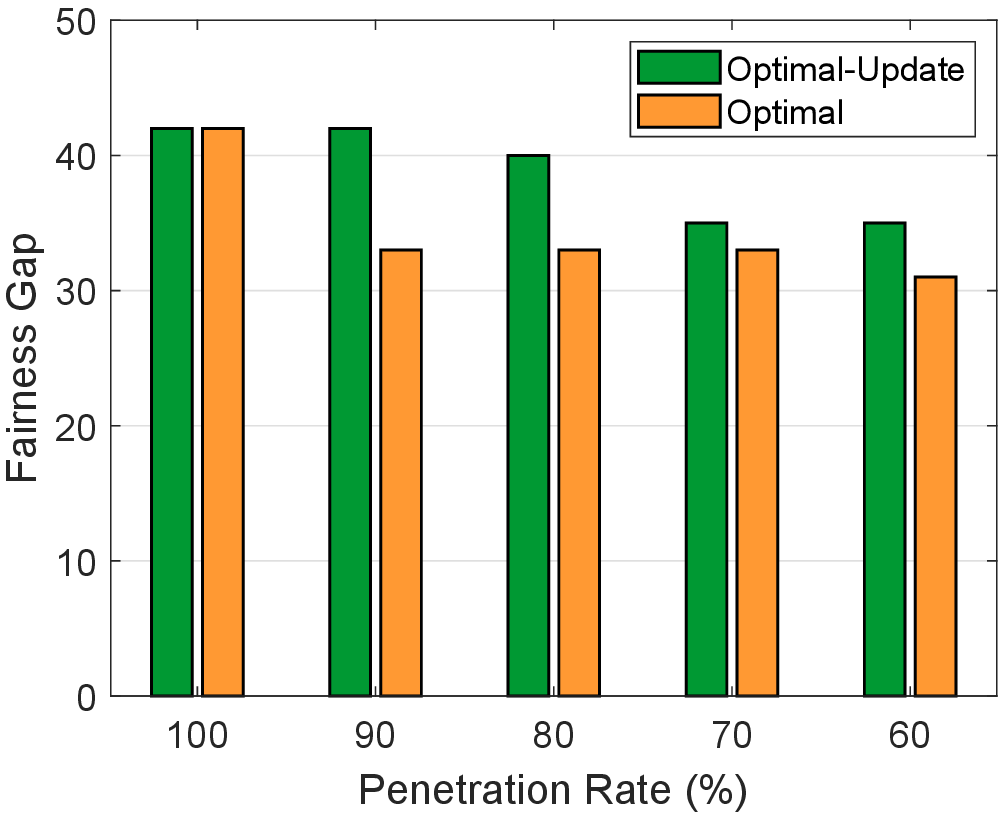}
\vspace{-15pt}
\caption {Effect of penetration rate on fairness after recomputing the communication schedule.}
\label{fig:PenetrationVsFairnessAdvanced}
\end{minipage}
\end{figure}

A salient feature of LOC-LPWAN is that the communication schedule between sensors and vehicles can be updated when a vehicle scheduled to relay packets does not show up. As such, if there are other vehicles available to participate, other vehicles can be selected to relay the packets instead. Therefore, in this experiment, we assume that the communication schedule is continually updated to locate and allow backup vehicles to participate. We first measure the throughput by varying the penetration rate under this assumption and compare the results with that obtained without this assumption. Fig.~\ref{fig:PenetrationVsThroughputAdvanced} shows the results. Interestingly, we observe that by rerunning and utilizing backup vehicles, LOC-LPWAN is able to maintain higher throughput by 8.5\% on average compared with the scenario where LOC-LPWAN does not update the communication schedule continually. We perform the similar experiment with respect to the degree of fairness. Results are shown in Fig.~\ref{fig:PenetrationVsFairnessAdvanced}. An interesting observation is that updating the communication schedule does not guarantee increased degree of fairness. More specifically, we observe that as new participating vehicles are selected with the updated schedule, it is possible that the fairness gap can be increased when the newly selected vehicles relay packets more than the maximum number of packets (or smaller than the minimum number of packets).

\subsection{Execution Time}
\label{sec:execution_time}

\begin{figure}[h]
\centering
\includegraphics[width=.9\columnwidth]{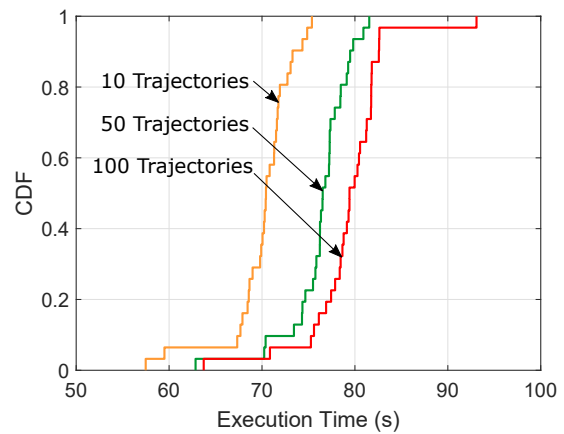}
\caption {The cumulative distribution function (CDF) graph of the execution time of LOC-LPWAN for varying number of trajectories.}
\label{fig:execution_time}
\end{figure}

A key feature of LOC-LPWAN is the capability of updating the communication schedule adaptively in response to dynamically changing traffic conditions. Therefore, the execution time of LOC-LPWAN should be fast. In this section, we measure the execution time of LOC-LPWAN with different numbers of trajectories, \emph{i.e.,} 10, 50, and 100. The results are depicted in Fig.~\ref{fig:execution_time}. Overall, LOC-LPWAN computes the optimal communication schedule quickly. Specifically, the average execution time is 70 sec, 71.8 sec, and 76.2 sec, for 10, 50, and 100 trajectories, respectively. An interesting observation is the large fixed runtime regardless of the sample size. To understand the reason for this large fixed runtime, we use the Matlab profiler~\cite{profiler} to identify which functions take the longest time. We find that there are two functions corresponding to constraint (5) and constraint (11) that take significantly more time than all other parts of the execution. More precisely, the implementation of these functions require the pre-allocation of a large array and operations on them regardless of the number of trajectories. If we add the runtime of these two functions and the optimizer, that corresponds to roughly 87\% of the total runtime. On the other hand, the computation time of the greedy algorithm is extremely fast at the cost of suboptimal throughput, fairness, and delay; specifically, the average execution time for the greedy algorithm is 0.26 sec, 0.32 sec, 0.37 sec, for 10, 50, and 100 trajectories, respectively.

\subsection{Summary of Findings}
\label{subsec:Summary_of_Findings}

In this section, we summarize the key findings of the simulation study. The throughput of LOC-LPWAN significantly higher than the greedy and greedy-N algorithms by up to 146.7\%. Compared with the baseline approach, with the same amount of budget, LOC-LPWAN achieves nearly 2X higher throughput, indicating about 50\% of cost savings. In terms of the degree of fairness, LOC-LPWAN outperforms both the greedy and greedy-N algorithms by decreasing the fairness gap by 65.7\% and 63.9\%, respectively. Furthermore, we demonstrate that LOC-LPWAN clearly meets the delay bound constraint.

\section{Discussions}
\label{sec:discussion}

In this section, we discuss several possible limitations of LOC-LPWAN and how those limitations can be addressed.

\subsection{Expected Vehicles Not in Service}
\label{subsec:vehicles_not_in_service}

It may happen that an expected vehicle is not be in service and thus the packets supposed to be relayed by the vehicle can be lost. LOC-LPWAN handles such situations by continually updating the communication schedule (\emph{i.e.,} rerunning the algorithm with updated $\mathbf{R_v}$, $\mathbf{Tr_v}$) to account for dynamically changing traffic conditions. Therefore, even if a vehicle is found to be not in service, other vehicle can be selected to forward the packets instead.

\subsection{Cost for Non-Subscription-Based LPWANs}
\label{subsec:cost_for_lpwans_with_no_subscription}

There are two types of cost models for LPWAN, \emph{i.e.,} the subscription-driven (SD) and manufacturing-driven (MD) networks~\cite{buurman2020low}. Most LPWANs are SD networks while there some MD networks such as LoraWAN~\cite{buurman2020low}. While the SD networks incur subscription fees, the MD networks require the operation and maintenance cost because the MD networks require the establishment and frequent monitoring of the direct communication links between LPWAN nodes and the gateway. Especially in urban areas, since the communication range of LPWAN nodes is relatively limited (\emph{e.g.,} 5 km~\cite{buurman2020low}) and the channel condition fluctuates, the MD networks require very careful node deployment plans and frequent channel status monitoring, which costs a high amount of maintenance fees.

The objective of LOC-LPWAN is to reduce the cost regardless of the cost model for LPWANs. Specifically, the subscription fees as well as the maintenance cost can be saved significantly by delivering packets via vehicles exploiting the high mobility and wide availability of vehicles. We expect that the C-V2X service will be available based on a monthly or yearly subscription as we have already seen that numerous automotive companies such as Tesla~\cite{tesla}, GM~\cite{gm}, and Toyota~\cite{toyota} have been providing the cellular service based on a monthly or yearly subscription. Furthermore, the PC5 connection, \emph{i.e.,} direct vehicle to vehicle communication may not rely on the cellular resource directly. The direct communication can be used to facilitate communication between vehicles and nearby infrastructure~\cite{5g1}\cite{5g2}. Besides C-V2X, there are other easily available options for enabling vehicles to relay the packets to save the cost. For example, a vehicle may defer forwarding a packet until it detects a free Wi-Fi coverage relays the packet via the Wi-Fi network which is becoming more and more widely available especially in the urban areas. Additionally, the driver's smartphone can also be used to relay the packets.




\subsection{Distribution of Communication Schedule}
\label{sec:}

Once the server finishes computing the communication schedule, the scheduling information is sent to sensors and vehicles so that the sensors can transmit packets to the vehicles according to the schedule. An interesting question is how the scheduling information is sent to the sensors. If the regular LPWAN channel is used, an extra cost will be incurred especially when the communication schedule is update frequently to account for dynamically changing traffic conditions. To address this issue, we adopt a receiver-initiated communication approach such as~\cite{fahmida2020long}. The idea is to allow the server to send the scheduling information to vehicles instead of sensors using the communication channel that vehicles are already subscribed to, \emph{e.g.,} C-V2I, a cellular network via driver's smartphone, or even home Wi-Fi (before a vehicle starts traveling). On the other hand, sensors operate in a low-power listening mode such as~\cite{fahmida2020long} that performs clear channel assessment (CCA) for a short period of time with extremely low power consumption to detect the channel activity, and when a channel activity is detected, it wakes up and receives packets transmitted from vehicles. This way sensors passively listen to the common channel and simply receives a packet when a vehicle transmits a packet according to the communication schedule, thereby eliminating the needs for the sensors to receive the scheduling information from the server.

\subsection{Adaptability of LOC-LPWAN}
\label{subsec:adaptability}

Sensor data collection is a multi-dimensional issue; as such, the performance of LOC-LPWAN can be enhanced by integrating with advanced communication technologies and path planning algorithms. Especially, it should be noted that LOC-LPWAN can be easily integrated with advanced communication and path planning algorithms to achieve better performance. More specifically, a better communication module/algorithm can be adopted by modeling the proposed framework with a finer timescale $T = \{t_1, t_2, ..., t_n\}$, \emph{i.e.,} the matrix $\matr{M}^{rx}$ can be defined on a finer timescale to model more effective sensor to vehicle communication. Additionally, while LOC-LPWAN makes a scheduling decision based on the user-provided vehicle trajectories $Tr_v^{i} = \{t_{v,i}^1, t_{v,i}^2, ..., t_{v,i}^{|T_v^{i}|}\}$, it can be reconfigured to suggest better paths by incorporating existing path planning algorithms to further improve the performance. Furthermore, it is also worth to note that LOC-LOPWAN can achieve evenly and timely data transmission even if vehicles are not present because the basic operation of LOC-LPWAN is to utilize the regular LPWAN channel while using vehicles as much as possible to reduce the operation cost.

\section{Conclusion}
\label{sec:conclusion}

We have presented LOC-LPWAN, a novel optimization framework designed to reduce the operation cost of roadside LPWAN sensors by using the cross platform technology. LOC-LPWAN allows sensors to offload sensor data to passing vehicles that forward the data to the LPWAN server, thereby reducing the cost for data transmissions. An optimization framework is developed to find the optimal communication schedule between sensors and vehicles to maximize the total amount of data transmitted given the maximum budget and minimum amount of compensation for each participating vehicle. The optimization framework also minimizes the gap between the largest and smallest number of data units transmitted by sensors to maximize the fairness by preventing sensors in certain areas from dominating data transmissions. It also ensures that all data units are transmitted with a specific delay bound. Numerical analysis performed with real-world trajectories of taxis demonstrates that LOC-LPWAN has significantly higher throughput compared with a greedy algorithm. Additionally, the fairness of data transmissions is substantially increased in LOC-LPWAN, and all data units are transmitted exactly within the specified time bound.

We expect that LOC-LPWAN will be a very useful tool for academia, industry, and government agencies who plan to develop ITS applications that deploy a large number of roadside LPWAN sensors and operate the network for a long period of time to save the cost significantly. While LOC-LPWAN is specifically designed for roadside sensors, it can be easily applied to other applications. For example, sensors deployed in a farm will be able to communicate with the smartphones of farmers, tractors, and UAVs to save the cost. Also, the cost for sensors deployed in a building for a smart building application can be saved by applying our work to allow the sensors to communicate with the smartphones of people in the building.


\bibliographystyle{IEEEtran}
\bibliography{mybibfile}

\vskip 10pt plus -1fil

\begin{IEEEbiography}[{\includegraphics[width=1in,height=1.25in,clip,keepaspectratio]{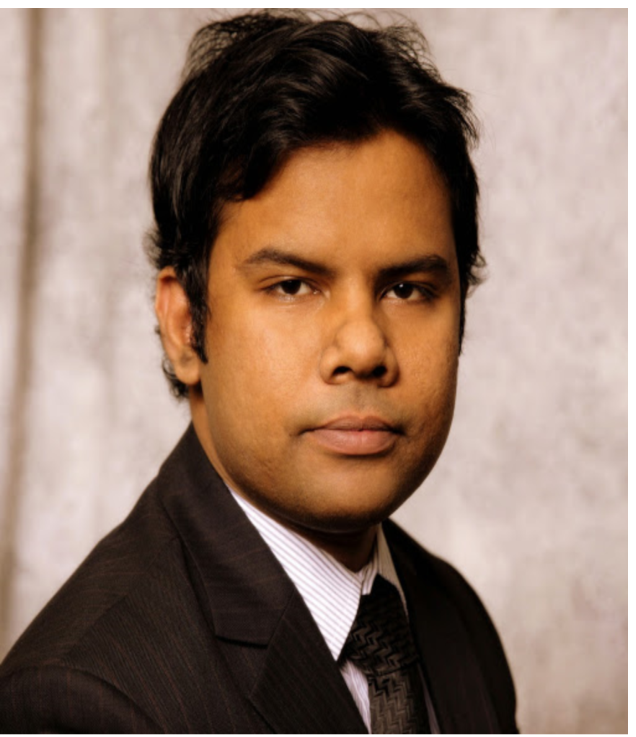}}]{Navid Imran}
received a B.Sc. degree in Electrical Engineering from South Dakota State University in 2016. Currently he is a Ph.D. student in the Department of Computer Science at the University of Memphis. His current research interests include intelligent transportation systems and wireless communication especially 5G, Vehicle-to-everything (V2X) communication and their optimization and security.
\end{IEEEbiography}

\vskip 10pt plus -1fil

\begin{IEEEbiography}[{\includegraphics[width=1in,height=1.25in,clip,keepaspectratio]{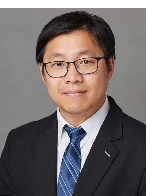}}]{Myounggyu Won}
(M'13) received a Ph.D. degree in Computer Science from Texas A\&M University at College Station, in 2013. He is an Assistant Professor in the Department of Computer Science at the University of Memphis, Memphis, TN, United States. Prior to joining the University of Memphis, he was an Assistant Professor in the Department of Electrical Engineering and Computer Science at the South Dakota State University, Brookings, SD, United States from Aug. 2015 to Aug. 2018, and he was a postdoctoral researcher in the Department of Information and Communication Engineering at Daegu Gyeongbuk Institute of Science and Technology (DGIST), South Korea from July 2013 to July 2014.  His research interests include smart sensor systems, connected vehicles, mobile computing, wireless sensor networks, and intelligent transportation systems. He received the Graduate Research Excellence Award from the Department of Computer Science and Engineering at Texas A\&M University - College Station in 2012.
\end{IEEEbiography}

\end{document}